\documentclass[fleqn,twoside]{article}
\usepackage{espcrc2}

\usepackage{graphicx}
\usepackage[figuresright]{rotating}

\newcommand{\url}[1]{{\tt #1}}
\newcommand{\lsim}
{\;\raisebox{-.3em}{$\stackrel{\displaystyle <}{\sim}$}\;}

\title{Prediction for the Lightest Higgs Boson Mass in the CMSSM using
  Indirect Experimental Constraints} 

\author
    {
      O.~Buchmueller\address[CERN]
      {
	CERN, CH-1211 Geneve 23, Switzerland
      },
      R.~Cavanaugh\address[Florida] 
      {
	Physics Department, University of Florida, Gainesville, Florida
        32611-8440, U.S.A. 
      },
      A.~De Roeck\addressmark[CERN]\hbox{$^{\rm ,}$}\address[Antwerpen]{
      Universitaire Instelling Antwerpen, B-2610 Wilrijk, Belgium
      },
      S.~Heinemeyer\address[Santander]
      {
	Instituto de Fisica de Cantabria (CSIC-UC), Santander, Spain
      },
      G.~Isidori\address[Isidori]
      {
	INFN, Laboratori Nazionali di Frascati, Via E. Fermi 40, 
           I-00044 Frascati, Italy
      }, 
      P.~Paradisi\address[Paradisi]
      {
	Departament de F\'{\i}sica Te\`orica  and IFIC, 
        Universitat de Val\`encia--CSIC, E--46100 Burjassot, Spain
      },
      F.J.~Ronga\addressmark[CERN],
      A.M.~Weber\address[MPI]
      {
	Max Planck Inst. f\"ur Phys., Foehringer Ring 6, D-80805 Munich,
        Germany 
      },
      G.~Weiglein\address[Durham]
      {
	IPPP, University of Durham, Durham DH1 3LE, U.K.
      }
    }
       
\begin{document}

\begin{abstract}
Measurements at low energies provide interesting indirect information about masses of particles that are (so far) too heavy to be produced directly.  Motivated by recent progress in consistently and rigorously calculating electroweak precision observables and flavour related observables, we derive the {\em preferred} value for $m_{\rm h}$ in the Constrained Minimal Supersymmetric Standard Model (CMSSM), obtained from a fit taking into account electroweak precision data, flavour physics observables and the abundance of Cold Dark Matter. No restriction is imposed on $m_{\rm h}$ itself: the experimental bound from direct Higgs boson search at LEP is not included in the fit.  A multi-parameter $\chi^2$ is minimized with respect to the free parameters of the CMSSM, $M_0$, $M_{1/2}$, $A_0$, $\tan\beta$.
A statistical comparison with the Standard Model fit to the electroweak precision data is made.
The preferred value for the lightest Higgs boson mass in the CMSSM is found to be $m_{\rm h}^{\rm CMSSM} = 110^{+8}_{-10}~{\rm (exp.)} \pm 3~{\rm (theo.)~GeV}/c^{2}$, where the first uncertainty is experimental and the second uncertainty is theoretical. This value is compatible with the limit from direct Higgs boson search at LEP.

\bigskip
\noindent IPPP/07/43 \hfill DCPT/07/86

\noindent MPP-2007-142
\vspace{-0.5cm}
\end{abstract}

\maketitle

\section{Introduction}

Low-energy supersymmetry (SUSY) is a prom\-is\-ing candidate for new physics beyond the Standard Model (SM). The Minimal Supersymmetric extension of the Standard Model (MSSM) has all SM multiplets extended to SUSY multiplets. The Higgs sector of the MSSM with two scalar doublets accommodates five physical Higgs bosons: the light and heavy CP-even h and H, the CP-odd A, and the charged Higgs bosons H${}^\pm$. In the MSSM no specific assumptions are made about the underlying SUSY-breaking mechanism, and a parametrization of all possible soft SUSY-breaking terms is used, introducing more than 100 new parameters in addition to those of the SM. While in principle these parameters could be independent of each other, experimental constraints from flavour-changing neutral currents, electric dipole moments, etc.\ seem to favour a certain degree of universality among the soft SUSY-breaking parameters. More precisely, present data favour models where the breaking of flavour universality is induced only by the Yukawa interaction, as in the general Minimal Flavour Violating scenario~\cite{D'Ambrosio:2002ex}. The assumption that the soft SUSY-breaking parameters are completely flavour blind at some high input scale, before renormalization, is frequently employed to further reduce the number of free parameters. The model focussed on in this paper, based on this simplification, is the Constrained MSSM (CMSSM), in which all the soft SUSY-breaking scalar masses are assumed to be universal ($M_0$) at the Grand Unified Theory (GUT) scale, as are the soft SUSY-breaking gaugino masses ($M_{1/2}$) and trilinear couplings ($A_0$). Additional parameters at the electroweak scale are (after imposing correct electroweak symmetry breaking) $\tan\beta$, the ratio of the two vacuum expectation values, and the sign of the Higgs mixing parameter, $\mu$.  All other parameters at the electroweak scale, including the Higgs boson masses, can be obtained from the CMSSM parameters by the help of renomalization group equations (RGE).
This very minimal model is sometimes referred to as minimal super-gravity (mSUGRA).

Measurements at low energies provide interesting indirect information about masses of particles that are (so far) too heavy to be produced directly. It is well known~\cite{Appelquist:1974tg,Dobado:1997up} that predicting the masses of SUSY particles using low-energy precision data is more difficult than it was for the top-quark mass due to the decoupling of the heavy sparticles. Nevertheless, several early analyses~\cite{deBoer:1996vq,deBoer:1996hd,Cho:1999km,Cho:2001nf,Erler:1998ur,Altarelli:2001wx} involving precision data have been performed in the context of the unconstrained MSSM.  More recently, many studies have been performed to extract the preferred values for the CMSSM parameters using low-energy precision data, bounds from astrophysical observables and flavour related observables~\cite{Djouadi:2001yk,deBoer:2001nu,deBoer:2003xm,Belanger:2004ag,Ellis:2003si,Ellis:2004tc,Ellis:2005tu,Ellis:2006ix,Ellis:2007aa,Baltz:2004aw,Allanach:2005kz,Allanach:2006jc,Allanach:2006cc,Allanach:2007qk,deAustri:2006pe,Roszkowski:2006mi}. These analyses differ in the precision observables that have been considered, the level of sophistication of the theory predictions that have been used and the way the statistical analysis has been performed. This latter point is developed below.  Motivated by recent progress in consistently and rigorously calculating electroweak precision observables~\cite{Degrassi:2002fi,Heinemeyer:2004gx,Heinemeyer:2006px,Z0Weber} and by new numerical studies of flavour related observables~\cite{Isidori:2006pk,Isidori:2007jw,Lunghi:2006uf} in the context of the MSSM, this study combines the most recent results from these observables in the framework of the CMSSM and extends the work presented in Ref.~\cite{yellowbook}. 

One of the most important predictions of the MSSM is the existence of a light neutral Higgs boson, with an upper bound $m_{\rm h} \lsim 135$~GeV$/c^{2}$~\cite{Degrassi:2002fi,Heinemeyer:1998np} (including loop corrections). This bound sensitively depends on $m_{\rm t}$, with $\delta m_{\rm h}/\delta m_{\rm t} \approx 1$~\cite{Heinemeyer:1999zf}. It incorporates parametric uncertainties from the experimental errors of $m_{\rm t}$ and the other input parameters, as well as uncertainties from unknown higher-order corrections. The upper bound on $m_{\rm h}$ for $m_{\rm t} = 170.9$~GeV$/c^{2}$, neglecting theoretical uncertainties, is about $129$~GeV$/c^{2}$~\cite{Degrassi:2002fi,Heinemeyer:1998np,Heinemeyer:1998yj}. Within the CMSSM, due to the lack of freedom to arrange all soft SUSY-breaking parameters independently, the upper bound is reduced by about 8~GeV$/c^{2}$ to $\sim127$~GeV$/c^{2}$~\cite{Heinemeyer:2004gx,Dedes:2003cg} (for scalar top masses not much larger than a few TeV$/c^{2}$). On the other hand, the direct search for a Higgs boson at LEP~\cite{Barate:2003sz,Schael:2006cr} (and to a lesser extent at the Tevatron~\cite{Abazov:2006hn,Abulencia:2006aj,Bernardi:2006fd,Abazov:2006ih,Abulencia:2005kq}) already imposes strong limits on the SM and (C)MSSM parameter space. 

Within the SM, the fit of the Higgs boson mass obtained from precision data yields~\cite{LEPEWWG}: 
\begin{eqnarray}
m_{\rm H}^{\rm SM} = 76^{+33}_{-24}~{\rm GeV}/c^{2},
\label{MH_SM}
\end{eqnarray}
with an upper limit of 144~GeV$/c^{2}$ at 95\%~C.L. 
Here, the recently obtained lower value of $m_{\rm t}$~\cite{:2007bx} plays an important role, and increases
the tension with the direct experimental limit obtained at LEP~\cite{Barate:2003sz}:
\begin{eqnarray}
m_{\rm h} > 114.4~\textrm{GeV}/c^{2}\textrm{ at 95\%\ C.L.}
\label{MH_LIMIT}
\end{eqnarray}
The corresponding bound within the MSSM can be substantially lower due to a reduced ZZh coupling or due to different, more complicated decay modes of the Higgs bosons~\cite{Schael:2006cr}. It has been shown~\cite{Ellis:2004tc,Ambrosanio:2001xb}, however, that, within the CMSSM, these mechanisms cannot be realised and, consequently, the experimental lower bound of 114.4~GeV$/c^{2}$ can be applied. This limit leaves only a very small part of the parameter space
unexcluded, taking into account the theoretical upper bound of $\sim 127$~GeV$/c^{2}$.

The aim of this paper is to derive the {\em preferred} value for $m_{\rm h}$ in the CMSSM, from a fit taking into account electroweak precision data, flavour physics observables and the abundance of Cold Dark Matter (CDM)~\cite{Spergel:2006hy}. {\em No} restrictions on $m_{\rm h}$ itself are imposed, {\it i.e.} the experimental bound from direct Higgs boson search at LEP is left out of the fit. A multi-parameter fit is performed by scanning the free parameters of the CMSSM, $M_0$, $M_{1/2}$, $A_0$ and $\tan\beta$, as well as several other SM parameters, including the top-quark mass $m_{\rm t}$. In order to comply with the anomalous magnetic moment of the muon, only positive values of~$\mu$ are considered~\cite{Bennett:2006fi,Moroi:1995yh}.

Indirect determinations of the lightest MSSM Higgs boson mass using precision data have been performed in the context of the CMSSM in the literature~\cite{Djouadi:2001yk,Ellis:2004tc,Ellis:2005tu,Ellis:2006ix,Ellis:2007aa,Allanach:2005kz,Allanach:2006jc,Allanach:2006cc,Allanach:2007qk,deAustri:2006pe,Roszkowski:2006mi,Dedes:2003cg,Ambrosanio:2001xb,Ellis:2002gp}. In most cases, however, the direct search bound from LEP was {\it a priori} included~\cite{Allanach:2006cc,Allanach:2007qk,Roszkowski:2006mi}. In Ref.~\cite{Ellis:2007aa}, preferred $m_{\rm h}$ values have already been obtained, yielding best-fit values of $m_{\rm h}$ close to $113-115$~GeV$/c^{2}$, depending on $\tan\beta$. Reference~\cite{Ellis:2007aa} used all relevant observables (potentially) sensitive to SUSY corrections and performed the analysis with and without taking into account the LEP bound on $m_{\rm h}$.
In that CMSSM scan, however, certain parameters were fixed, {\it e.g.}\ $\tan\beta = 10, 50$. 
In the present work instead, the full CMSSM parameter space is scanned and the $\chi^2$ is minimised with respect to all parameters, without any restrictions or constraints on $m_{\rm h}$. This procedure facilitates a comparison of the CMSSM prediction with the SM prediction and allows one to use the goodness-of-fit of fit probabilities to discuss a possible experimental preference of the precision data for the CMSSM or the SM hypothesis.

\section{Multi-parameter Fit to Experimental Observables}
\label{sec:mpfit}

\begin{table*}[!tbh]
\renewcommand{\arraystretch}{1.2}
\begin{center}
\begin{tabular}{|c|c|c|c|c|} \hline
Observable & Th. Source & Ex. Source & Constraint & Add. Th. Unc. \\ \hline \hline
$\Delta\alpha_{\rm had}^{(5)}(m_{\rm Z})$ 
                       &\cite{POPE} &\cite{ADLOS:2005em} &$0.02758\pm0.00035$ & -- \\
\hline
$ m_{\rm Z}$ [GeV$/c^{2}$]         &\cite{POPE}   &\cite{ADLOS:2005em} &$91.1875\pm0.0021$  & -- \\ 
\hline
$ \Gamma_{\rm Z}$ [GeV$/c^{2}$]    &\cite{POPE}   &\cite{ADLOS:2005em} &$2.4952\pm0.0023$   & 0.001  \\ 
\hline
$\sigma_{\rm had}^{0}$ [nb] 
                       &\cite{POPE} &\cite{ADLOS:2005em} &$41.540\pm0.037$    & -- \\
\hline
$R_l                 $ &\cite{POPE}   &\cite{ADLOS:2005em} &$20.767\pm0.025$    & -- \\ 
\hline
$ A_{\rm fb}(\ell)          $ &\cite{POPE}   &\cite{ADLOS:2005em} &$0.01714\pm0.00095$ & -- \\ 
\hline
$ A_{\ell}(P_\tau)      $ &\cite{POPE}   &\cite{ADLOS:2005em} & 0.1465 $\pm$ 0.0032 & -- \\ 
\hline
$ R_{\rm b}                $ &\cite{POPE}   &\cite{ADLOS:2005em} & 0.21629 $\pm$ 0.00066 & -- \\ 
\hline
$ R_{\rm c}                $ &\cite{POPE}   &\cite{ADLOS:2005em} & 0.1721 $\pm$ 0.003 & -- \\ 
\hline
$ A_{\rm fb}({\rm b})          $ &\cite{POPE}   &\cite{ADLOS:2005em} & 0.0992 $\pm$ 0.0016 & -- \\ 
\hline
$ A_{\rm fb}({\rm c})          $ &\cite{POPE}   &\cite{ADLOS:2005em} & 0.0707 $\pm$ 0.0035 & -- \\ 
\hline
$ A_{\rm b}                $ &\cite{POPE}   &\cite{ADLOS:2005em} & 0.923 $\pm$ 0.020 & -- \\ 
\hline
$ A_{\rm c}                $ &\cite{POPE}   &\cite{ADLOS:2005em} & 0.670 $\pm$ 0.027 & -- \\ 
\hline
$ A_\ell({\rm SLD})           $ &\cite{POPE}   &\cite{ADLOS:2005em} & 0.1513 $\pm$ 0.0021 & -- \\ 
\hline
$ \sin^2 \theta_{\rm w}^{\ell}(Q_{\rm fb})$ 
                       &\cite{POPE}   &\cite{ADLOS:2005em} & 0.2324 $\pm$ 0.0012 & -- \\ 
\hline
$ m_{\rm W}          $ [GeV$/c^{2}$] &\cite{POPE}   &\cite{ADLOS:2005em} & $80.398\pm0.025$ & 0.010 \\ 
\hline
$ m_{\rm t}          $ [GeV$/c^{2}$] &\cite{POPE}   &\cite{ADLOS:2005em} & $170.9\pm1.8$    & -- \\ 
\hline
$ BR_{\rm b \to s \gamma}^{\rm SUSY}/ BR_{\rm b \to s \gamma}^{\rm SM} $
                       &\cite{bsg_note} &\cite{hfag}       & 1.13 $\pm$ 0.12 & 0.15 \\ 
\hline
$ BR_{\rm B_{s} \to \mu^{+} \mu^{-}} $ &\cite{Belanger:2006is} &\cite{hfag}&$<8.0\times 10^{-8}$ & $0.02\times10^{-8}$ \\ 
\hline
$ a_{\mu}^{\rm SUSY} - a_{\mu}^{\rm SM}$ 
                       &\cite{Moroi:1995yh}    &\cite{Bennett:2006fi,Davier:2007ua,Hertzog:2007hz} 
                                                           &$(29.5 \pm 8.7)\times10^{-10}$ &$2.0\times10^{-10}$ \\ 
\hline
$ \Omega h^2         $ &\cite{Belanger:2006is,Belanger:2001fz,Belanger:2004yn} 
                                               &\cite{Spergel:2006hy} &$0.113\pm0.009$ & 0.012 \\ 
\hline
$ m_{\rm h}          $ [GeV$/c^{2}$] & \cite{Degrassi:2002fi,Heinemeyer:1998np,Heinemeyer:1998yj,Frank:2006yh} 
                                                   & \cite{Barate:2003sz} & see text & see text  \\ 
\hline
\end{tabular}
\caption{List of experimental constraints used in this work. The values and errors shown are the current best understanding of these constraints. The rightmost column displays additional theoretical uncertainties taken into account when implementing these constraints in the CMSSM. The constraint on $m_{\rm h}$ is only used in the first part of this study.\label{tab:constraints}} 
\end{center}
\end{table*}

The observables taken into account in the fit are listed in Table~\ref{tab:constraints}. The RGE running from the GUT to the electroweak scale is performed with the help of the program {\tt SoftSusy}~\cite{Allanach:2001kg}. At the electroweak scale, the calculations of flavour physics observables are based on Refs.~\cite{Isidori:2006pk,Belanger:2006is}, electroweak precision observables on Refs.~\cite{Heinemeyer:2006px,Z0Weber,Moroi:1995yh}, Higgs boson observables on {\tt FeynHiggs}~\cite{Degrassi:2002fi,Heinemeyer:1998np,Heinemeyer:1998yj,Frank:2006yh}, and the CDM density on {\tt micrOMEGAs}~\cite{Belanger:2006is,Belanger:2001fz,Belanger:2004yn}. 
All calculations are combined by a single steering code~\cite{yellowbook}, which takes advantage of the SUSY Les Houches Accord~\cite{Skands:2003cj} to ensure consistency of the input parameters. 

Based on these observables, a global $\chi^2$ function is defined, which combines all calculations with experimental constraints: 
\[
\chi^2 = \sum_i^{N}\frac{(C_i - P_i)^2}{\sigma(C_i)^2 + \sigma(P_i)^2} +
\sum_i^{M} \frac{(f_{{\rm SM}_i}^{\rm obs} - f_{{\rm SM}_i}^{\rm
    fit})^2}{\sigma(f_{{\rm SM}_i})^2}  
\label{eqn:chi2}
\]
Here $N$ is the number of observables studied, each $C_i$ represents an experimentally measured value (constraint) and each $P_i$ defines a CMSSM parameter-dependent prediction for the corresponding constraint.  Each predicted CMSSM parameter set $\{P_i\}$ is checked to ensure that none of the LEP experimental limits on sparticle masses are violated~\cite{LEP_SUSY_WG_1,LEP_SUSY_WG_2,LEP_SUSY_WG_3,LEP_SUSY_WG_4}. The $M$ standard model parameters $f_{\rm SM}=\{\Delta\alpha_{\rm had}(m_{\rm Z}), m_{\rm t}, m_{\rm Z}\}$ are included as fit parameters and constrained to be within their current experimental resolution $\sigma(f_{\rm SM})$.
This procedure ensures that the uncertainties of the most relevant standard model parameters are properly included in the multi-parameter $\chi^2$ fit. Particular care has been taken to ensure that all theoretically predicted observables, $P_i$, are consistently defined and calculated at loop level~\cite{yellowbook}. In the case where the constraint on $m_{\rm h}$ is applied (Eq.~\ref{MH_LIMIT}), its experimental error is parameterized from the $\chi^{2}$ distribution of the measurement~\cite{Barate:2003sz}, and a theoretical uncertainty of 3~GeV$/c^{2}$ is set on its prediction~\cite{Degrassi:2002fi,Heinemeyer:2004gx}.

A fit is then performed to determine the compatibility of a given set of CMSSM parameters with the experimental constraints defined in Table~\ref{tab:constraints}. The minimization of the $\chi^2$ is carried out by initially sampling the parameter space with Monte Carlo ``pseudo-experiments.''  With each pseudo-experiment the $\chi^2$ is determined by minimizing over all free parameters using the package {\tt Minuit}~\cite{James:1975dr}.  Once a multi-dimensional region of interest is identified, {\tt Minuit} is used to precisely locate the $\chi^2$ minima.   

This work builds upon previous studies but differs from them in several respects.  First, in Ref.~\cite{Ellis:2007aa}, $\tan\beta$ was fixed to 10,~50, $A_0$ was varied as $0$, $\pm 1$, $\pm 2~M_{1/2}$, and $M_0$ was fixed to yield the correct amount of CDM. A two-dimensional $\chi^2$ scan over $M_{1/2}$ and $A_0$ was then performed, and provided information about preferred regions in the CMSSM parameter space.  In the study presented here instead,
all free parameters are placed in the overall $\chi^2$ minimum by the fit, thus removing the need to fix any model parameters during the scans.  Indeed, in the present work, only experimental constraints are imposed when deriving confidence level contours, without any direct constraints on model parameters themselves. 
Hence, the results presented here have a clearer statistical meaning and are more general with respect to previous studies.

\begin{figure*}[!tbh]
\begin{picture}(150,180) 
  \put(0,-10){ \resizebox{7.7cm}{!}{\includegraphics{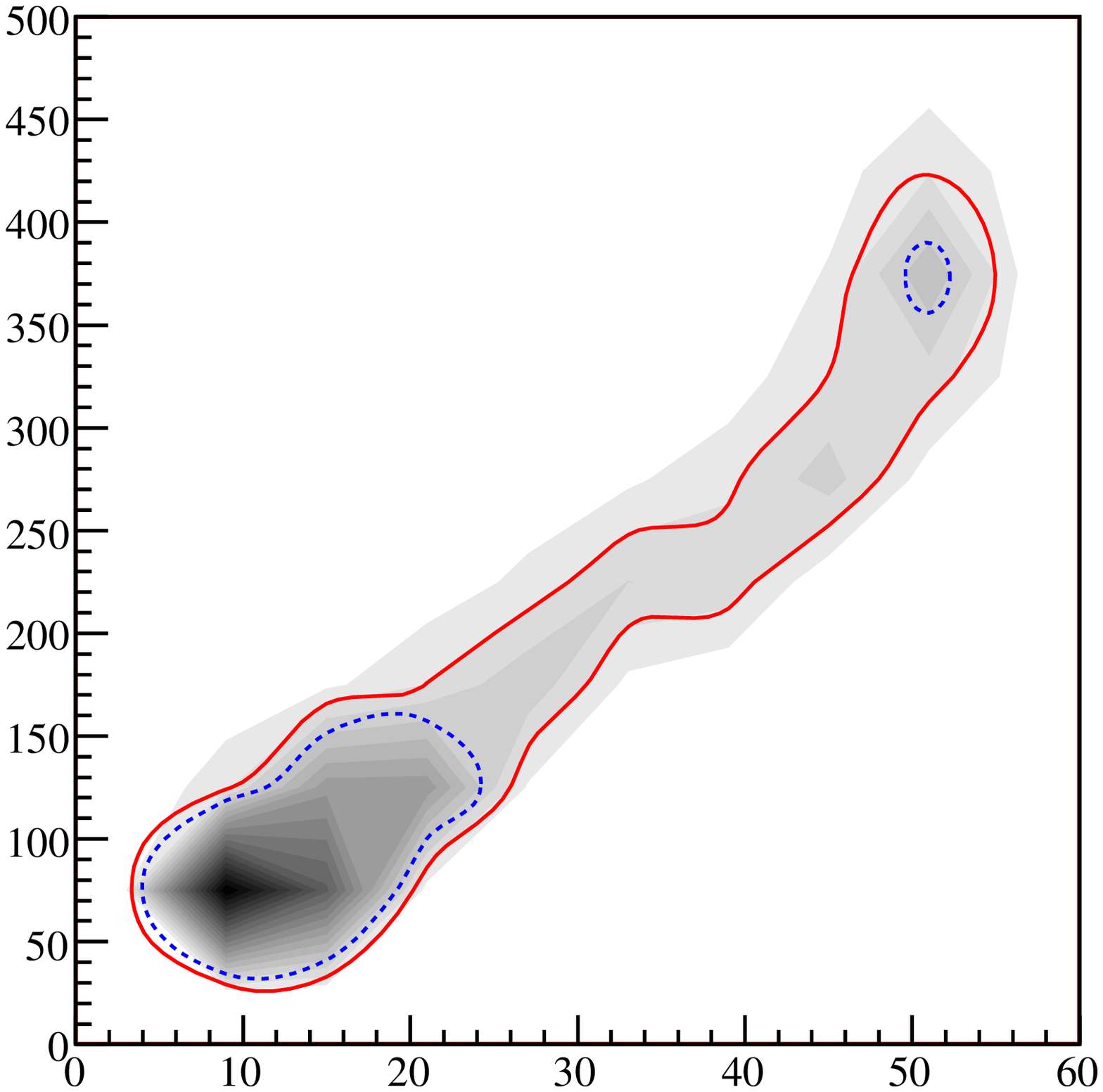}}  }
  \put(175, -10){$\tan \beta$}
  \put(10, 125){\begin{rotate}{90}$M_0$ [GeV$/c^{2}$]\end{rotate}}
  \put(220,-10){ \resizebox{7.7cm}{!}{\includegraphics{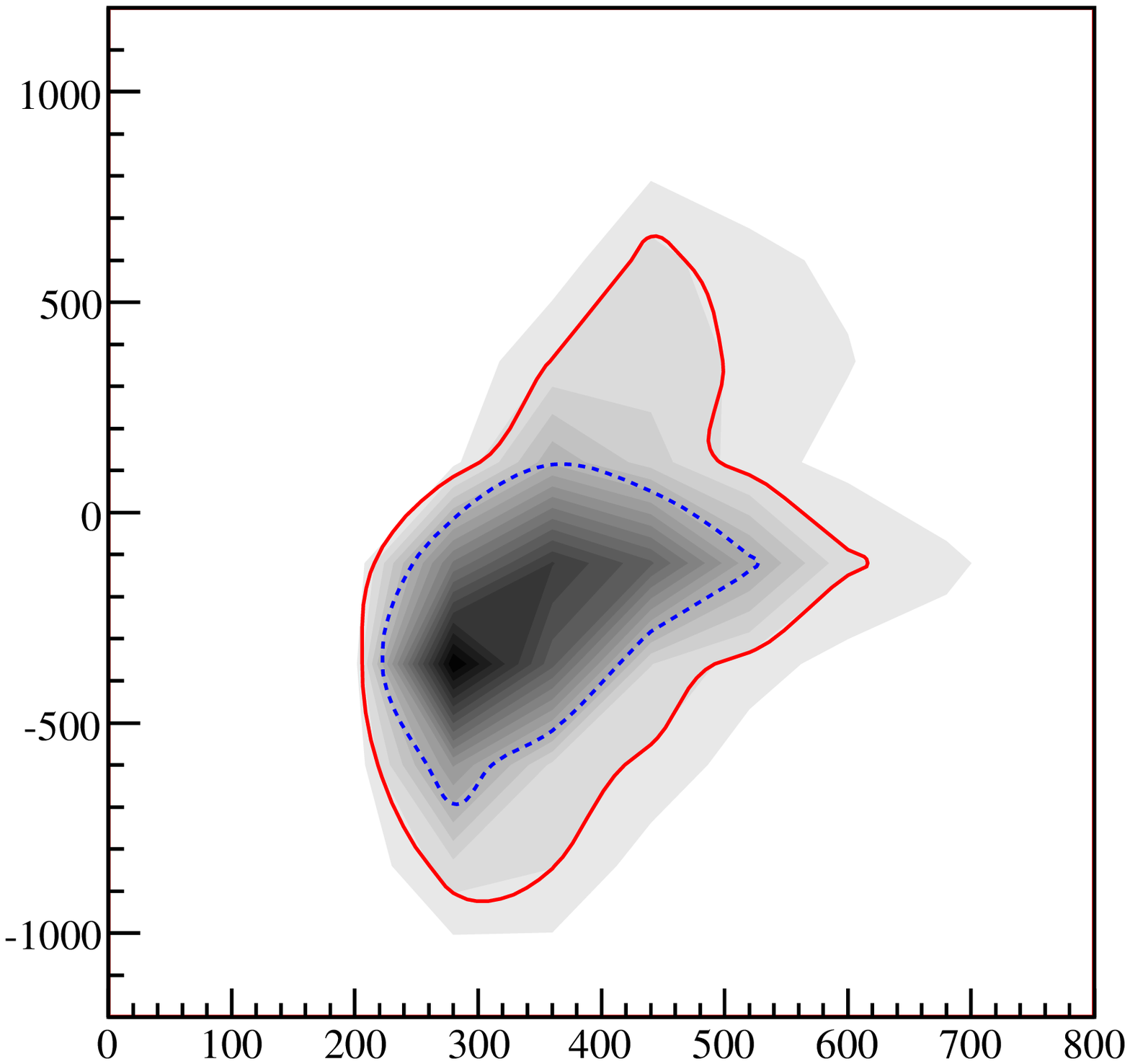}}  }
  \put(370, -10){$M_{1/2}$ [GeV$/c^{2}$]}
  \put(230, 125){\begin{rotate}{90}$A_0$ [GeV/$c^{2}$]\end{rotate}}
\end{picture}
\caption {Left:  Two parameter contour in the ($\tan \beta$, $M_0$)
  plane with 68\% (dotted) and 95\% (solid) confidence level regions.
  The two dotted regions together represent 68\% total probability.
  Right:  Two parameter contour in the ($M_{1/2}$, $A_0$) plane with
  68\% (dotted) and 95\% (solid) confidence level regions. Darker shading corresponds
  to higher confidence level.} 
\label{fig:contours}
\end{figure*}

Second, in Ref.~\cite{Ellis:2003si} a likelihood analysis of the CMSSM parameter space was performed, but $m_{\rm h}$ was not emphasized.
Third, in Refs.~\cite{Baltz:2004aw,Allanach:2005kz,Allanach:2006jc,Allanach:2006cc,Allanach:2007qk,deAustri:2006pe}, Markov Chain Monte Carlo techniques were employed to sample the entire CMSSM parameter space with respect to the likelihoods and the Bayesian posterior probabilities. The resulting probabilty distributions are usually graphically displayed in two-dimensional planes by integrating over the unseen dimensions. Given the limited experimental precision of the data, these probability maps are somewhat dependent on the subjective prejudice that went into the original parameterization of the priors on the CMSSM parameters (see also the discussion in Ref.~\cite{Allanach:2007qk}). Although the Bayesian analyses in \cite{Baltz:2004aw,Allanach:2005kz,Allanach:2006jc,Allanach:2006cc,Allanach:2007qk,deAustri:2006pe} provide interesting information on what to expect at future colliders, the prior dependence can be avoided by the use of a purely $\chi^2$ based fit as done in Ref.~\cite{Ellis:2007aa}.  Further, the $\chi^2$ probability, $P(\chi^2,N_{\rm dof})$, properly accounts for the number of degrees of freedom, $N_{\rm dof}$, and thus represents a quantitative measure for the quality-of-fit.  Hence $P(\chi^2,N_{\rm dof})$ can be used to estimate the absolute probability with which the CMSSM describes the experimental data.
 In the present study, $P(\chi^2,N_{\rm dof})$ is found to have a flat distribution using Monte Carlo pseudo-experiments, thus yielding a reliable estimate of the confidence level.  The use of pseudo-experiments has another advantage in that no assumptions of Gaussian behaviour have to be made.  This leads to a robust estimate of the 68\% and 95\% confidence level contours where, for multiple separated contours of the same probability, individual probabilities are added in order to obtain the desired total probability. This property of the frequentist approach was also recently exploited in Ref.~\cite{Allanach:2007qk} by using profile likelihoods.

  In addition, unlike the present work, the analyses of Refs.~\cite{Baltz:2004aw,Allanach:2005kz,Allanach:2006jc,Allanach:2006cc,Allanach:2007qk,deAustri:2006pe} apply the direct experimental search results from LEP when estimating the most probable value for $m_{\rm h}$.

Finally, by using the complete set of flavour and electroweak observables listed in Table~\ref{tab:constraints}, this work exploits additional experimental information compared to other studies. (A very similar set has, however, been used in Ref.~\cite{Ellis:2007aa}.) 

\section{Results}

\begin{figure}[!hbt]
\begin{picture}(250,200) 
  \put(0,-15){ \resizebox{7.8cm}{!}
               {\includegraphics{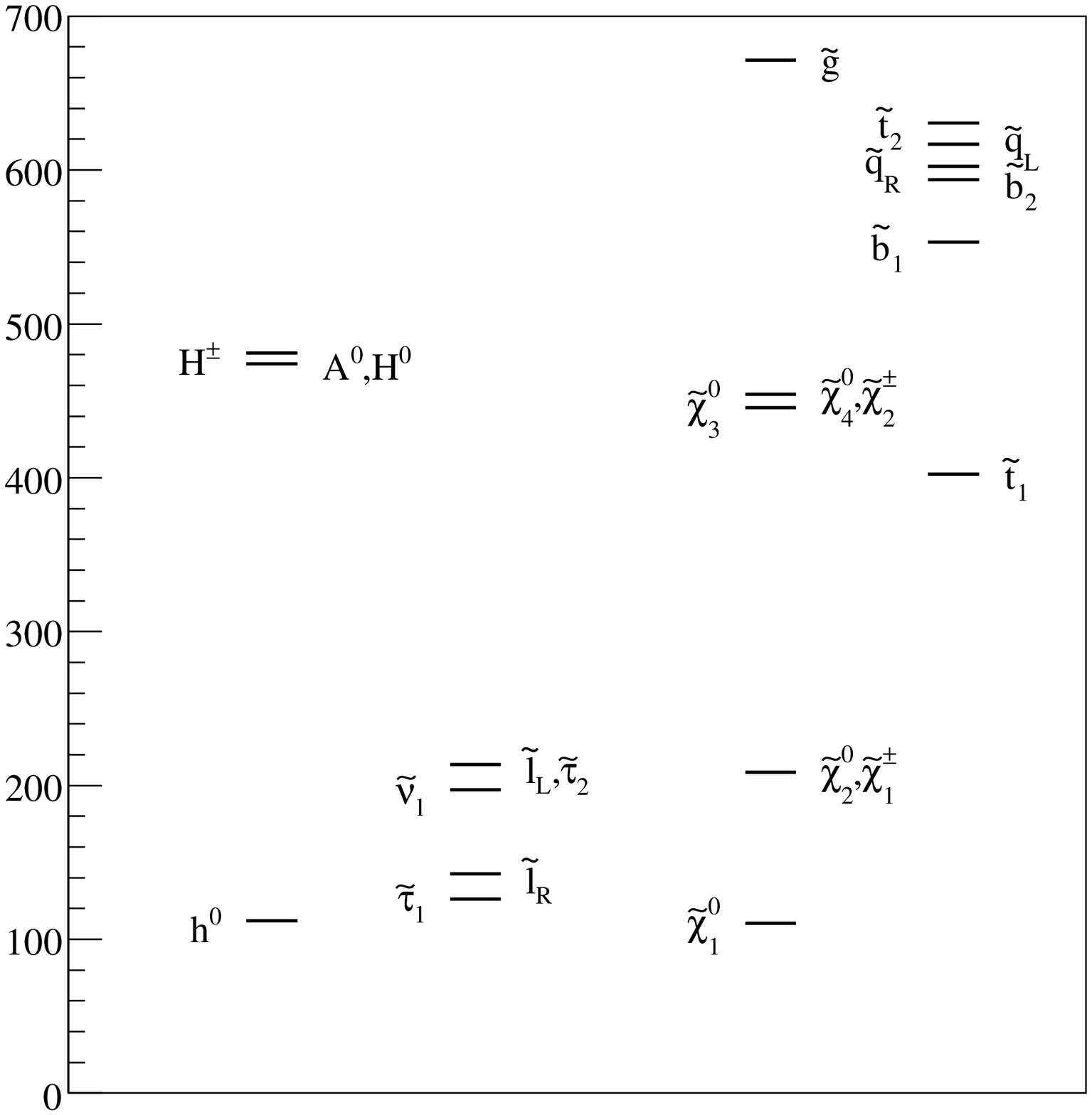}}  }
  \put(5, 125){\begin{rotate}{90}Mass [GeV$/c^{2}$]\end{rotate}}
\end{picture}
\caption{Mass spectrum of super-symmetric particles at the globally preferred 
         $\chi^{2}$ minimum.
         Particles with mass difference smaller than 5~GeV/$c^{2}$ have been
         grouped together.}
\label{fig:spectrum}
\end{figure}

Using the $\chi^2$ function defined in Section \ref{sec:mpfit}, the CMSSM parameter space is explored. The regions of the CMSSM parameter space that are still consistent with all existing data, including the bound on $m_{\rm h}$ from direct LEP searches (Eq.~\ref{MH_LIMIT}) but neglecting the Tevatron bounds on the Higgs sector, are first focussed on. In order to map these regions, contours involving $\tan\beta$, $M_0$, $M_{1/2}$, and $A_0$ are shown in Fig.~\ref{fig:contours}.  The left plot of Fig.~\ref{fig:contours} displays a two parameter contour in the ($\tan \beta$, $M_0$) plane and illustrates the 68\% (dotted) and 95\% (solid) confidence level regions. A corridor with two distinct minima is observed along the diagonal.  The globally preferred minimum corresponds to small $\tan \beta$ and small $M_0$, while the second, less preferred minimum ($\Delta\chi^2\approx 1.8$ between the two minima) corresponds to large $\tan \beta$ and relatively larger $M_0$. For completeness, the right plot of Fig.~\ref{fig:contours} displays the contours in the ($M_{1/2}$, $A_0$) plane. In addition to LEP experimental limits that are included in the fit, the sparticle spectrum is checked to be compatible with the latest limits from searches at the Tevatron~\cite{TevatronSearches}. The CMSSM parameters at the globally preferred minimum are listed in Table~\ref{tab:parameters}, together with their 1-sigma error. The corresponding sparticle mass spectrum is shown in Fig.~\ref{fig:spectrum}, just for illustrative purposes. 

\begin{table}[!hb]
\begin{center}
\renewcommand{\arraystretch}{1.2}
\begin{tabular}{|c|c|}
\hline
CMSSM parameter & Preferred value \\
\hline
\hline
$M_{0}$     & $ (85^{+40}_{-25})$~GeV/$c^{2}$\\
$M_{1/2}$   & $(280^{+140}_{-30})$~GeV/$c^{2}$\\
$A_{0}$     & $(-360^{+300}_{-140})$~GeV/$c^{2}$\\
$\tan\beta$ & $10^{+9}_{-4}$\\
sgn$(\mu)$  & $+1$ (fixed) \\
\hline
\end{tabular}
\caption{Values of the CMSSM parameters at the globally preferred $\chi^{2}$ minimum, and corresponding 1-sigma errors. The lower limit of Eq.~\ref{MH_LIMIT} is included.\label{tab:parameters}} 
\end{center}
\end{table}

Other studies \cite{Ellis:2007aa,Allanach:2006cc} have found qualitatively similar behaviour. Because this work uses a traditional $\chi^2$ fit, however, the $\chi^2$ probability can be used to estimate how well the CMSSM describes the experimental data.  At the global minimum, the CMSSM describes the experimental data rather well, giving a $\chi^2$ of 17.34 per 14 degrees of freedom, which corresponds to a fit probability of 24\%.   For comparison, the SM describes the same electroweak experimental data (with the LEP bound on $m_{\rm H}$ imposed; excluding the flavour physics observables, $a_\mu$ and the CDM constraint) with a $\chi^2$ of 19.4 per 14 degrees of freedom, or a fit probability of 15\% \cite{LEPEWWG}.

\begin{figure*}[!htb]
\begin{picture}(500,190) 
  \put(0,-10){ \resizebox{7.5cm}{!}
             {\includegraphics[angle=0]{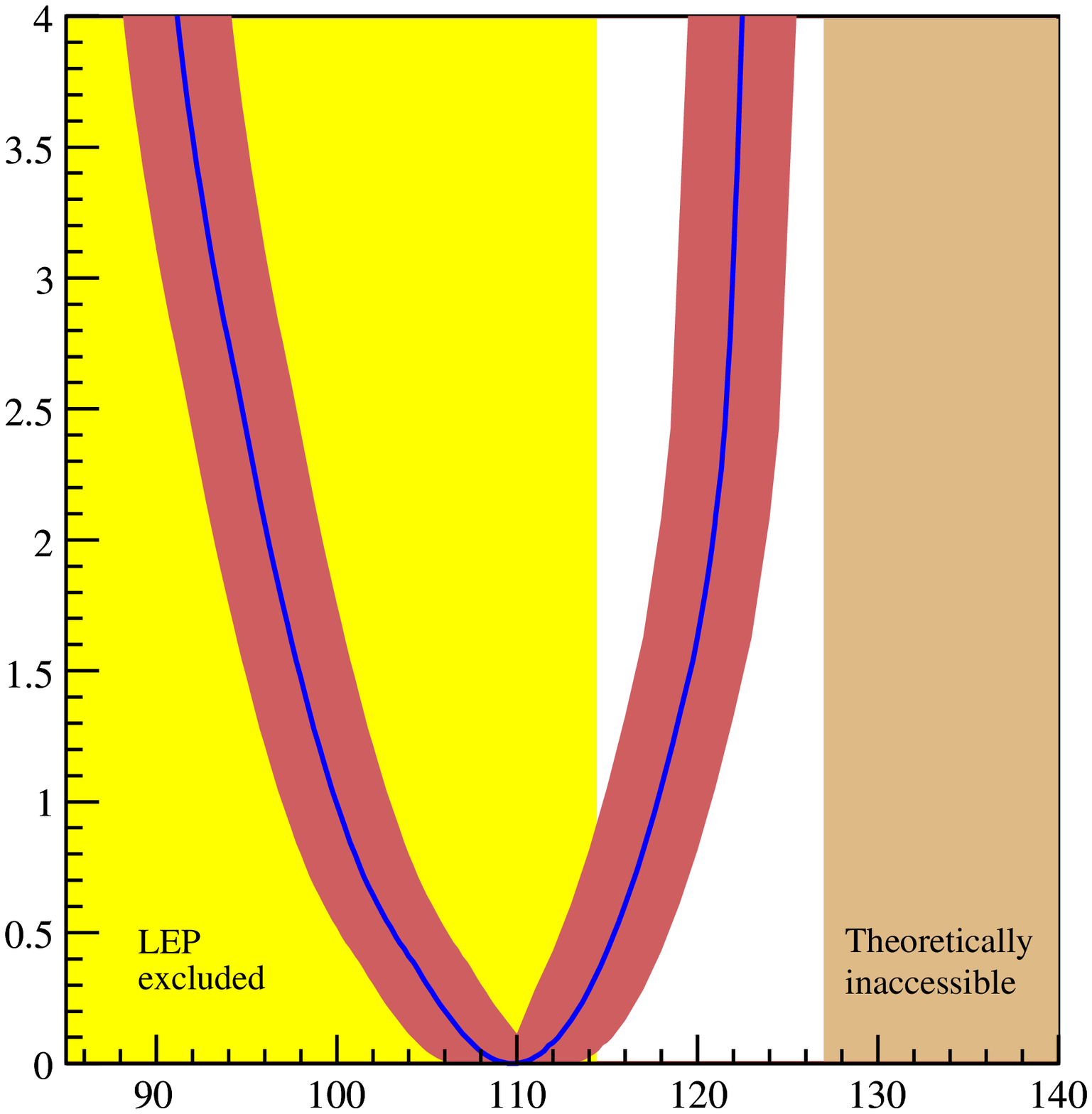}}  }
  \put(75, 160){CMSSM}
  \put(165, -10){$m_{\rm h}$ [GeV$/c^{2}$]}
  \put(10, 170){\begin{rotate}{90}$\Delta \chi^2$\end{rotate}}
  \put(220,-10){ \resizebox{7.5cm}{!}
               {\includegraphics{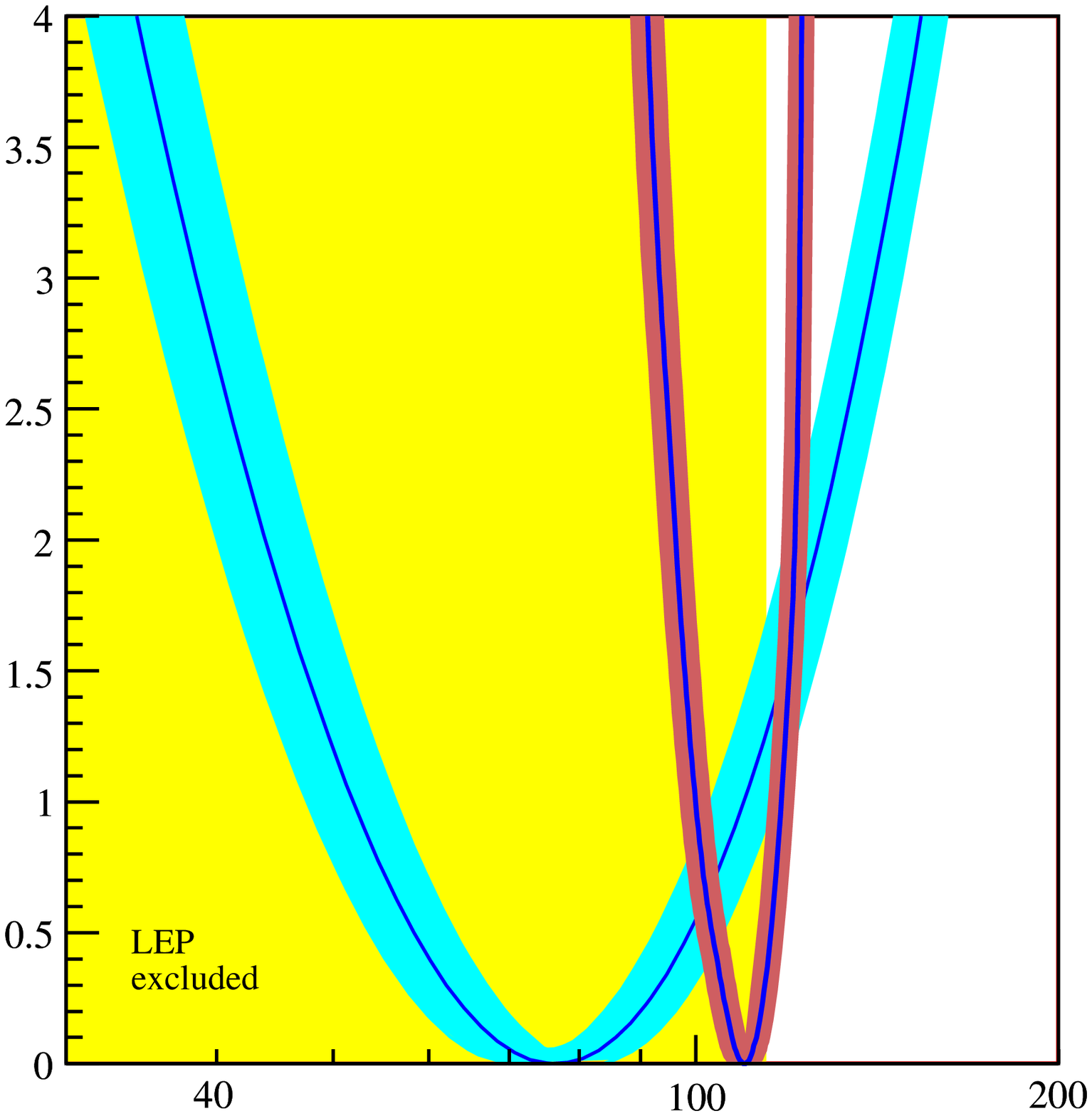}}  }
  \put(370, -10){$m_{\rm Higgs}$ [GeV$/c^{2}$]}
  \put(230, 170){\begin{rotate}{90}$\Delta \chi^2$\end{rotate}}
  \put(375, 40){CMSSM}
  \put(280, 160){SM}
\end{picture}
\caption{Left:  Scan of the lightest Higgs boson mass versus $\Delta \chi^2 =
  \chi^2 - \chi^2_{\rm min}$.  The curve is the result of a CMSSM fit
  using all of the available constraints listed in Table~\ref{tab:constraints}, 
  except the limit on $m_{\rm h}$.  
  The red (dark gray) band represents the total
  theoretical uncertainty from unknown higher-order corrections,
  and the dark shaded area on the right above 127 GeV$/c^{2}$ 
  is theoretically inaccessible (see text).
  Right:  Scan of the Higgs boson
  mass versus $\Delta \chi^2$ for the SM (blue/light gray), as
  determined by \cite{LEPEWWG} using all available electroweak
  constraints, and for comparison, with the CMSSM scan superimposed
  (red/dark gray). The blue band represents the total theoretical uncertainty
  on the SM fit from unknown higher-order corrections.} 
\label{fig:mh_vs_chi2}
\end{figure*}

Now we turn to the case where the bound on $m_{\rm h}$ from direct Higgs boson search at LEP are not incorporated and the {\em preferred} $m_{\rm h}$ values in the CMSSM can be derived.  The main result of this study is given as a one parameter scan in the lightest Higgs boson mass, presented in Fig.~\ref{fig:mh_vs_chi2}.
The $\chi^2$ is minimized with respect to all CMSSM parameters for each point of this scan. Therefore, $\Delta \chi^2=1$ represents the 68\% confidence level uncertainty on $m_{\rm h}$. 
Since the direct Higgs boson search limit from LEP is not used in this scan (unlike other studies~\cite{Allanach:2006cc,Roszkowski:2006mi}) the lower bound on $m_{\rm h}$ arises as a consequence of {\em indirect} constraints only. 

Several interesting features are worth noting. There is a well defined minimum, leading to a prediction of the light neutral Higgs boson mass of  
\begin{eqnarray}
m_{\rm h}^{\rm CMSSM} = 110^{+8}_{-10}~{\rm(exp.)} \pm 3~{\rm(theo.)~GeV}/c^{2}
\label{MH_CMSSM}
\end{eqnarray}
where the first, asymmetric uncertainties are experimental and the second uncertainty is theoretical (from the unknown higher-order corrections to $m_{\rm h}$~\cite{Degrassi:2002fi}).  The result obtained here is in good agreement with the previous results in Ref.~\cite{Ellis:2007aa}, where a simpler $\chi^2$~analysis has been performed.  The fact that the minimum in Fig.~\ref{fig:mh_vs_chi2} is sharply defined is a general consequence of the MSSM, where the neutral Higgs boson mass is not a free parameter. After including radiative corrections~\cite{Degrassi:2002fi,Heinemeyer:1998np,Chankowski:1992er,Dabelstein:1994hb,Heinemeyer:1998jw}, $m_{\rm h}$ is a well-defined function of the gauge couplings, $m_{\rm t}$, $m_{\rm Z}$ and soft SUSY-breaking parameters. The theoretical upper bound $m_{\rm h} \lsim 135 (127)$~GeV$/c^{2}$ in the (C)MSSM explains the sharper rise of the $\Delta\chi^2$ at large $m_{\rm h}$ values and the asymmetric uncertainty. In the SM, $m_{\rm H}$ is a free parameter and only enters (at leading order) logarithmically in the prediction of the precision observables. In the (C)MSSM this logarithmic dependence is still present, but in addition $m_{\rm h}$ depends on $m_{\rm t}$ and the SUSY parameters, mainly from the scalar top sector. The low-energy SUSY parameters in turn are all connected via RGEs to the GUT scale parameters. 
The sensitivity on $m_{\rm h}$ in the present analysis is therefore the combination of the indirect constraints on the four free CMSSM parameters and the fact that $m_{\rm h}$ is directly predicted in terms of these parameters.  
This sensitivity also gives rise to the fact that the fit result in the CMSSM is less affected by the uncertainties from unknown higher-order corrections in the predictions of the electroweak precision observables. While the theoretical uncertainty of the CMSSM fit (red/dark gray band in Fig.~\ref{fig:mh_vs_chi2}) is dominated by the higher-order uncertainties in the prediction for $m_{\rm h}$, the theoretical uncertainty of the SM fit (blue/light gray band in Fig.~\ref{fig:mh_vs_chi2}) is dominated by the higher-order uncertainties in the prediction for the effective weak mixing angle, $\sin^2\theta^{\rm eff}$~\cite{Awramik:2004ge}.

\begin{figure*}[!hbt]
\begin{picture}(500,250) 
  \put(-10,-15){ \resizebox{8.7cm}{!}
               {\includegraphics{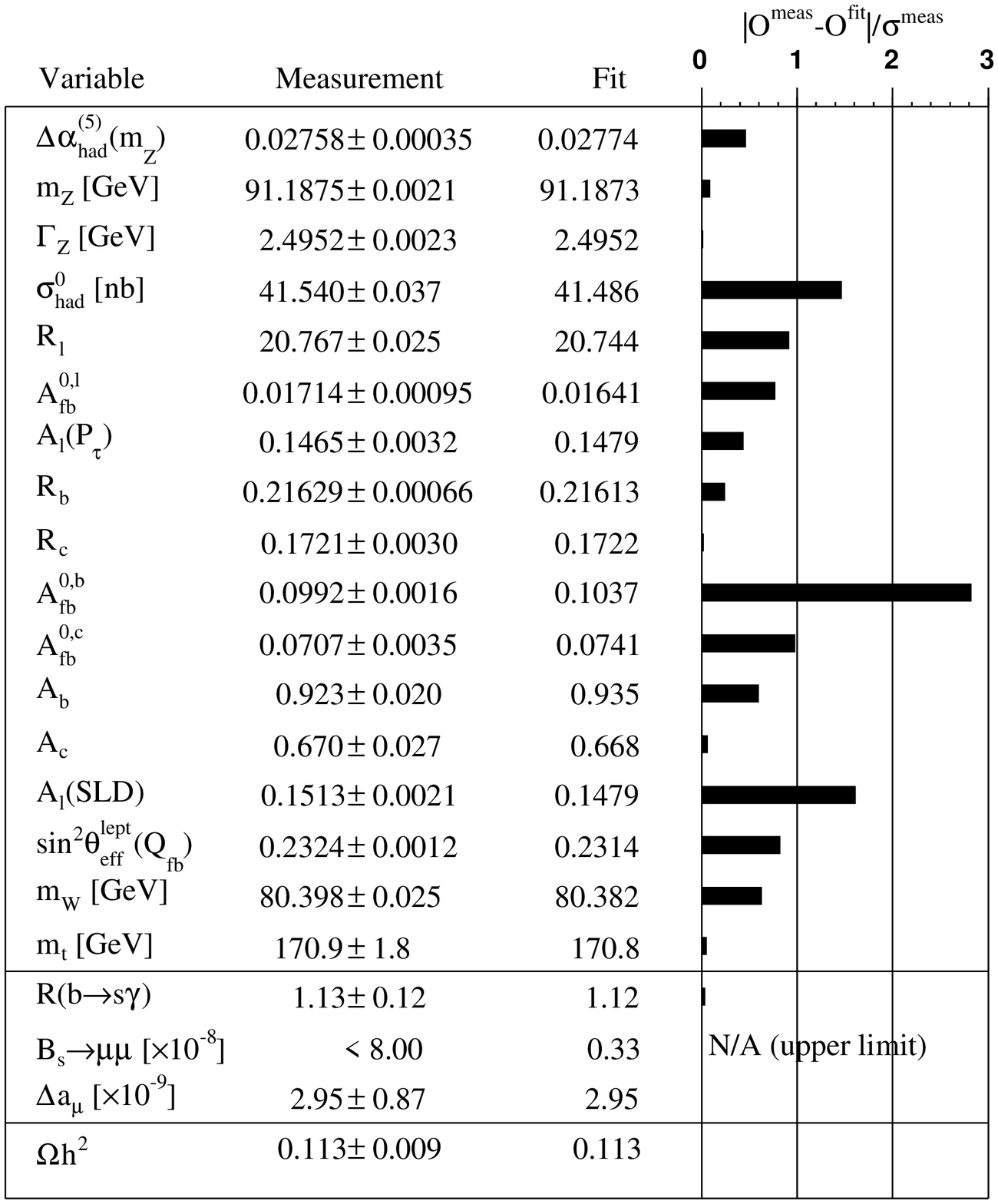}}  }
  \put(220,-15){ \resizebox{8.7cm}{!}
               {\includegraphics{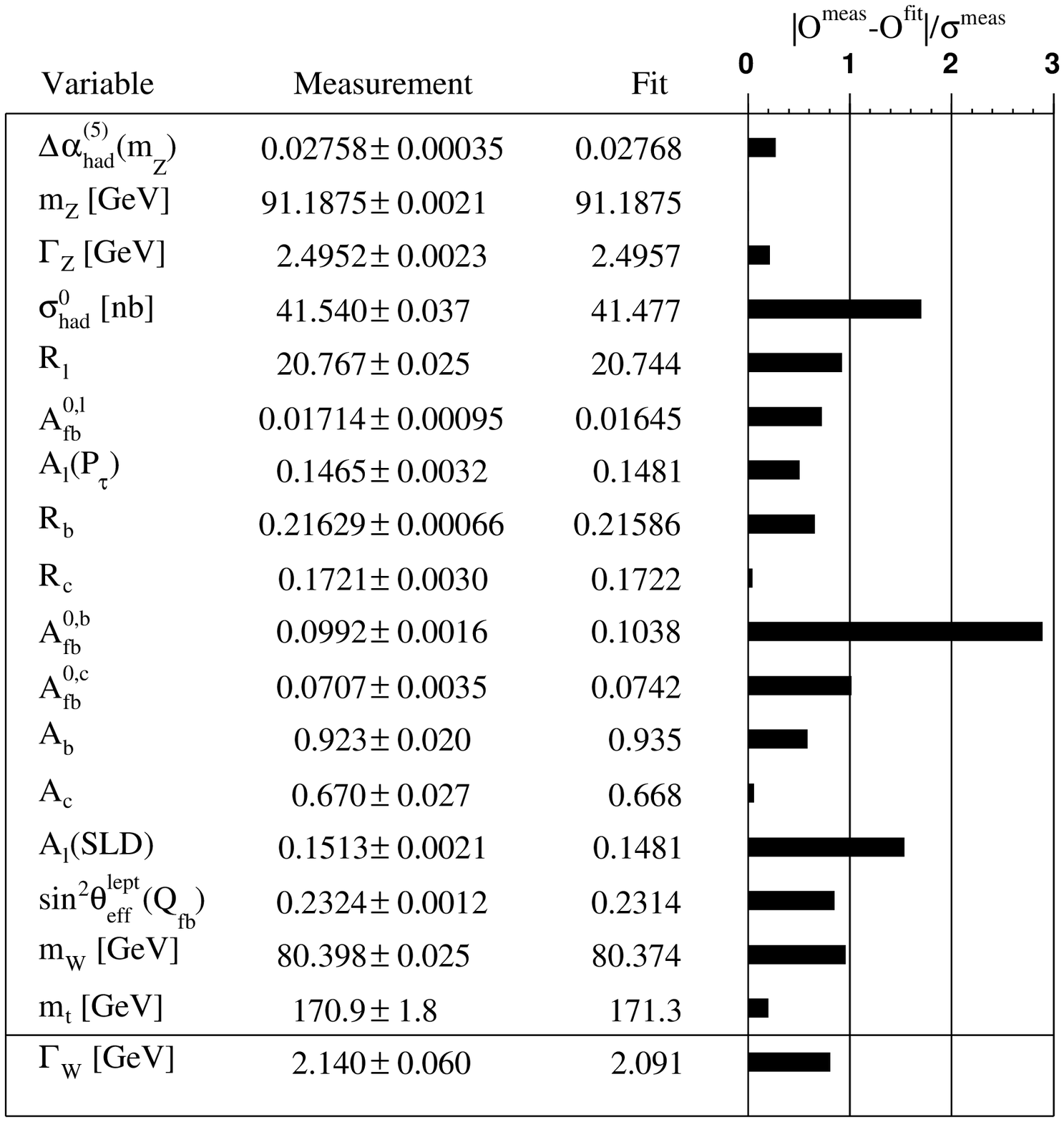}}  }
  \put(75,245){CMSSM}
  \put(280,245){Standard Model}
\end{picture}
\caption {Left:  Difference between the measured value and the fit value
  normalized by the measurement uncertainty, also known as ``pulls,''
  for all observables used in the CMSSM fit to experimental
  constraints. 
  Right:  Latest pulls for the SM as provided
  by \cite{LEPEWWG}. The lower limit of Eq.~\ref{MH_LIMIT} is not included
  in these fits.}
\label{fig:pulls}
\end{figure*}

The most striking feature is that even {\em without} the direct experimental lower limit from LEP of 114.4~GeV$/c^{2}$ (Eq.~\ref{MH_LIMIT}), the CMSSM prefers a Higgs boson mass which is quite close to and compatible with this bound. From the curve in Fig.~\ref{fig:mh_vs_chi2}, the value of the $\chi^{2}$ at the LEP limit corresponds to a probability of 20\% (including theoretical errors in the red band). This probability may be compared with the SM, 
where the indirect constraints on $m_{\rm H}$ implies $m_{\rm H}^{\rm SM} = 76^{+33}_{-24}$ {\rm GeV$/c^{2}$}, or a 12\% $\chi^{2}$ probability at the LEP limit (including theoretical errors from the blue band). While the tight mass range in the prediction of $m_{\rm h}^{\rm CMSSM}$ is a general expectation of the MSSM, the fact that the CMSSM prediction is in slightly better agreement with data than the SM prediction is a non-trivial result. The SM fits the experimental data reasonably well; however, the preferred value of its only free parameter ($m_{\rm H}$) implies a rather low Higgs boson mass. The CMSSM fits the same experimental data, supplemented by the flavour physics observables, $a_\mu$ and the CDM constraint, (Fig.~\ref{fig:pulls}) equally well (or slightly better) and the preferred values of its free parameters are such that the Higgs boson is predicted to be in better agreement with the Higgs boson searches at LEP.  
Interestingly enough, the CMSSM prediction is consistent with the possibility that the slight excess of Higgs-like events observed by LEP~\cite{Schael:2006cr,Barate:2000ts} could indeed stem from a SM-like Higgs boson. 
 
The pulls for the CMSSM, defined to be the difference between the measured value and the fit value normalized by the measurement uncertainty, are shown in the left plot of Fig.~\ref{fig:pulls} (still excluding $m_{\rm h}$ from the fit). 
They demonstrate that the CMSSM describes the data well, providing a $\chi^2$ of 17.0 per 13 degrees of freedom, or a 20\% goodness-of-fit probability.  This result may be compared with the pulls of the experimental observables used in a SM fit to electroweak data provided by \cite{LEPEWWG}, displayed in the right plot of Fig.~\ref{fig:pulls}.  The SM fit results in a $\chi^2$ of 18.2 per 13 degrees of freedom, or a 15\% goodness-of-fit probability~\cite{LEPEWWG}.

It should be noted that a key role in the determination of CMSSM parameters is played by the CDM constraints, $\rm b\to s \gamma$ and $a_\mu$. As shown in  Fig.~\ref{fig:pulls}, it is essentially impossible to distinguish between SM and CMSSM predictions in the electroweak precision observables.  
Indeed, because of the decoupling of virtual effects induced by sparticle loops, these observables provide mainly exclusion bounds on the sparticle spectrum. 
On the other hand, the three mentioned observables/constraints provide a first clue concerning the size of deviations from the SM (CDM cannot be explained in the SM, $a_\mu$ is in disagreement with the SM by more than 3$\,\sigma$ (using $e^+e^-$ input data for the hadronic vacuum polarization) and $\rm b \to s \gamma$ agrees reasonably well with the SM prediction), which is essential in constraining the CMSSM parameter space.  In particular, CDM constraints and $a_\mu$ are essential to extract upper bounds on $M_{1/2}$ and $M_0$ as a function of $\tan\beta$ and to fix the sign of $\mu$, while the addition of $\rm b\to s \gamma$ plays a key role in further constraining $M_{1/2}$ vs.~$\tan\beta$~\cite{Isidori:2006pk,Isidori:2007jw}. 

\section{Conclusion and outlook}

External constraints possess the potential to severely restrict new physics model parameters~\cite{Djouadi:2001yk,deBoer:2001nu,deBoer:2003xm,Belanger:2004ag,Ellis:2003si,Ellis:2004tc,Ellis:2005tu,Ellis:2006ix,Ellis:2007aa,Baltz:2004aw,Allanach:2005kz,Allanach:2006jc,Allanach:2006cc,Allanach:2007qk,deAustri:2006pe,Roszkowski:2006mi}.   However, when identifying regions of parameter space that are compatible with external constraints using high precision fits, it is important to rigorously enforce consistent definitions and predictions across all of the used experimental observables.  By including such considerations with care, a statistical analysis of the CMSSM has been performed, which allows all parameters to vary freely.  
Figure~\ref{fig:pulls} suggests that the CMSSM provides a good description, perhaps even slightly better than the SM, of the external experimental constraints used in this study.  
The use of a $\chi^{2}$ fit also allowed the result to be directly interpreted in terms of confidence levels. Also, in the context of CMSSM fits, for the first time, a {\em full} scan of the lightest MSSM Higgs boson mass has been performed, without incorporating the experimental bound from direct Higgs boson search at LEP.  The fit results can be compared with the scan of the SM Higgs boson, in the context of electroweak fits \cite{LEPEWWG}.  Without taking into account the existing LEP limit on the Higgs boson mass, the current indirect constraints on the CMSSM lead to a preferred range of the lightest Higgs boson mass of $110^{+8}_{-10}~{\rm(exp.)}~\pm 3~{\rm(theo.)}$~GeV$/c^{2}$, in agreement with previous analyses~\cite{Ellis:2007aa}. This value is significantly heavier than the SM prediction of $76^{+33}_{-24}$~GeV$/c^{2}$~\cite{LEPEWWG}. Finally the $\chi^2$ probabilities of the fits indicate that, without imposing the LEP experimental lower limit on the Higgs boson mass, both the CMSSM and SM describe the current experimental data reasonably well, but a slightly higher goodness-of-fit in the CMSSM hypothesis (20\% $\chi^{2}$ probability) compared with the SM (15\% $\chi^{2}$ probability) is observed.  If the LEP lower bound is imposed in the fit, the goodness-of-fit for the CMSSM fit increases to 24\% compared to the SM fit with 15\%.


Future improvements on the experimental~\cite{Erler:2000jg} and theoretical side should increase the sensitivity to new-physics parameters. Furthermore, similar studies in the framework of less restricted models, with more free parameters than the CMSSM, are foreseen~\cite{Ellis:2006ix,Ellis:2007aa}. In particular, the study of indirect constrained fits in the context of a reduced MSSM parameter set directly defined at the electroweak scale~\cite{yellowbook} can potentially provide important information on the SUSY Lagrangian (and complements analyses using future direct measurements~\cite{Lafaye:2004cn,Bechtle:2004pc}). Such studies have the advantage that the extracted parameters are defined at a scale similar to experimental observables, making the interpretation of potential new physics discoveries easier.

\section{Acknowledgements}
The authors gratefully thank Martin Gr\"unewald and Patrick Janot for many useful discussions related to this work. 
S.H.\ and G.W.\ thank John Ellis and Keith Olive for helpful discussions.
This work was supported in part by the European Community's Marie-Curie Research
Training Network under contracts MRTN-CT-2006-035505
`Tools and Precision Calculations for Physics Discoveries at Colliders'
and MRTN-CT-2006-035482 `FLAVIAnet', and by the Spanish MEC and FEDER under 
grant FPA2005-01678.



\end{document}